\documentclass[12pt]{article}
\usepackage{epsfig}

\newcommand{\be}{\begin{equation}}
\newcommand{\ee}{\end{equation}}
\newcommand{\ba}{\begin{eqnarray}}
\newcommand{\ea}{\end{eqnarray}}
\newcommand{\nn}{\nonumber}
\newcommand{\al}{\alpha}

\newcommand{\G}{\Gamma}

\newcommand{\MSsch}{{\overline{\rm MS}}}
\newcommand{\ice}[1]{\relax}

\begin{document}
\thispagestyle{empty}
\begin{flushright}
MZ-TH/99-65\\
December 1999\\
\end{flushright}
\vspace{0.5cm}
\begin{center}
{\Large\bf $t\bar t$ cross section near the 
production threshold in NNLO of NRQCD}\\[1truecm]
{\large \bf A.A.Pivovarov}

Institut f\"ur Physik, Johannes-Gutenberg-Universit\"at,\\
Staudinger Weg 7, D-55099 Mainz, Germany\\
and \\
Institute for Nuclear Research of the\\
Russian Academy of Sciences, Moscow 117312, Russia
\end{center}

\vspace{1truecm}
\begin{abstract}
The cross section near the $t\bar t$ production threshold 
in $e^+e^-$ annihilation and $\gamma\gamma$ collisions
is discussed.
The NNLO results within NRQCD are briefly overviewed.
\end{abstract}

\vskip 2.5cm
\begin{center}
{\it Prepared for Quarks-2000 Seminar, Pushkin, Russia\\

May 13-21, 2000}
\end{center}

\newpage
\noindent{\bf 1. Introduction.}

\noindent The hadron production near heavy quark threshold 
will be thoroughly studied experimentally at future accelerators,
e.g.~\cite{exp}. The motion of
the heavy quark-antiquark pair
near the production threshold is nonrelativistic to 
high accuracy that justifies the use 
of the nonrelativistic quantum mechanics as a proper 
theoretical framework for describing such a system 
\cite{CasLep1,CasLep2,CasLep3}. 
Being much simpler than the comprehensive relativistic treatment 
of the bound state problem
with Bethe-Salpeter amplitude \cite{BetheSal},
this approach allows one to take into account 
exactly such an essential feature
of the near-threshold dynamics as Coulomb interaction \cite{VL,ttee}.
For unstable heavy quarks with a large decay width it is possible to
compute the cross section near threshold point-wise in energy
because the large decay width  
serves as an infrared cutoff and suppresses the long distance
effects of strong interaction \cite{FK}.

\noindent{\bf 2. $t\bar t$-system near threshold} 

\noindent 
The $t\bar t$-pair near the production 
threshold is just a system that satisfies the requirement of being
nonrelativistic.
Therefore the description of $t\bar t$-system near the production
threshold $\sqrt{s}\approx 2m_t$ ($\sqrt{s}$ is a total energy
of the pair,
$m_t$ is the top quark mass) is quite precise within NRQCD.
Reasons for this accuracy are related to the large mass of the top: 
\begin{itemize}
\item
The top quark is very heavy $m_t=175~{\rm GeV}$ \cite{PDG}
and there is an energy region of about $8-10$ GeV near the 
threshold where the nonrelativistic approximation
for the kinematics is very precise.
The velocity is small (for $E$ of interest in the range 
$|E|\sim 2\div 5~{\rm GeV}$) 
\be
v=\sqrt{1-\frac{4m_t^2}{s}}=\sqrt{1-\frac{4m_t^2}{(2m_t+E)^2}}\simeq
\sqrt{\frac{E}{m_t}}\approx 0.10 \div 0.15 \ll 1 \, .
\ee
Relativistic effects are small and can be taken into account
perturbatively in $v$ (even in $v^2$).
\item
the strong coupling constant 
at the high energy scale is small 
$\al_s(m_t)\approx 0.1$ that makes 
the mapping of QCD onto the low energy 
effective theory (NRQCD), which is perturbative in $\al_s(m_t)$,
numerically precise.
\item
The decay width of top quark is large, $\G_t=1.43$ GeV;
infrared (small momenta) region is suppressed and PT calculation 
for the cross section near the threshold is reliable 
even point-wise in energy.
\end{itemize}
The above properties could make the $t\bar t$ system near threshold
flat and not interesting for physical study,
however, there is a feature of quark-antiquark 
interaction that becomes dominant in this kinematical regime 
and 
brings a nontrivial structure into the dynamics. 
Because $\al_s \sim v$ and the ratio $\al_s/v$ is not small,
the Coulomb interaction is enhanced.
The ordinary perturbation theory for 
the cross section (with free quarks as the lowest order
approximation)
breaks down and all terms of the order $(\al_s/v)^n$ should be summed.
The expansion for a generic observable $f(E)$ in this kinematical region
has the form
\be
\label{generic}
f(E)=f_0(\al_s/v)+\al_s f_1(\al_s/v)+\al_s^2 f_2(\al_s/v)+\ldots
\ee
where 
$f_i(\al_s/v)$ are some (not polynomial) functions of the ratio
$\al_s/v$, $f_0(\al_s/v)$ is a result of the pure Coulomb approximation
(or a kind of its improvement).
The expansion in $\al_s$ in eq.~(\ref{generic})
takes into account the perturbative QCD corrections to the parameters 
of NRQCD and relativistic corrections (in the regime $v\sim \al_s$).
One can study the $t\bar t$ system near 
threshold in the processes $e^+e^-\to t\bar t$ \cite{ttee,ttee1}
and $\gamma\gamma \to t\bar t$ \cite{ttgg,ttgg1}. These processes 
have the following features:
\begin{itemize}
\item
$e^+e^-\to t\bar t$: the production vertex is 
local (the electromagnetic current in case of photon
and/or weak current in case of $Z$-boson),
the basic observable is a production cross section
which is saturated by S-wave (for the vector current),
NNLO analysis is available.
\item
$\gamma\gamma \to t\bar t$: the production vertex is 
nonlocal (T-product of two electromagnetic currents),
both S- and P-waves can be studied for different helicity 
photons, the number of observables is larger 
(cross sections $\sigma_S$,
$\sigma_P$, S-P interference).
Because of nonlocality of the production vertex
the high energy coefficient (necessary for mapping the QCD quantities
to NRQCD ones) is more difficult to obtain. 
Some calculations were done in NLO \cite{KMgg,Piv}
and the full analysis of the cross section
is available in NLO of NRQCD only \cite{PP2}. 
The low energy part of the process can be studied in NNLO
without a strict normalization to full QCD (see \cite{INOK} for
relativistic corrections).
\end{itemize}

\noindent {\bf 3. Theoretical description.}

\noindent I shall discuss only $e^+e^-\to t\bar t$ process
mediated by the photon 
because NNLO analysis, which contains the most interesting features,
is possible.
The basic quantity is the vacuum polarization
function
\be
\Pi(E)=i \int \langle T j_{em}(x)j_{em}(0)\rangle e^{iqx} dx, 
\quad q^2=(2m_t+E)^2\, .
\ee
Near the threshold (for small energy $E$) NRQCD is used.
The cross section is saturated with 
S-wave scattering. In this approximation 
the polarization function near the threshold to the NNLO
accuracy in NRQCD is given by 
\be
\Pi(E)=\frac{2\pi}{m_t^2}C_h(\al_s)C_{\cal O}(E/m_t) G(E;0,0)\, .
\label{Rv}
\ee 
The pole mass definition is used for $m_t$ (e.g. \cite{Tar}),
$\al_s$
is the strong coupling constant. The choice of normalization points 
for coupling constants entering different parts of the theoretical
expression (\ref{Rv})
will be discussed later.
$C_h(\al_s)$ is the high energy coefficient
which has been
known in the NLO since long ago \cite{Kar,HarBr,Chi}
(before the explicit formulation of NRQCD). 
$G(E;0,0)$ is the nonrelativistic GF, $E=\sqrt{s}-2m_t$.
The leading term of the cross section in the effective theory
representation is given by a
correlator of currents with dimensionality 3 in mass units.
The quantity $C_{\cal O}(E/m_t)$ describes the contributions 
of higher dimension operators within the effective theory
approach. These contributions have, in general, a different
structure than the leading term.
However, to the NNLO of NRQCD the contribution
of higher dimension operators  
can be written as a total factor $C_{\cal O}(E/m_t)$
for the leading order GF,
$C_{\cal O}(E/m_t)=1-4E/3m_t$.
The polarization function near the threshold (\ref{Rv})
contains expansions in small parameters
$\al_s$ and/or $v$, cf. eq.~(\ref{generic}).
The leading order approximation 
of the low energy part is
the exact Coulomb solution for the Green's function.
The more detailed description of 
the ingredients of the representation in eq.~(\ref{Rv}) is:
\begin{itemize}
\item the nonrelativistic Green's function $G(E;x,x')$
is given by $G(E)=(H-E)^{-1}$ where 
\be
\label{hamNR}
H=\frac{p^2}{m_t}+V(r)
\ee
is the nonrelativistic Hamiltonian
describing dynamics of the $t\bar t$-pair near the threshold.
The most complicated part of Hamiltonian (\ref{hamNR})
to find is the heavy quark static
potential $\tilde V_{pot}(q)$ entering 
into the potential $V(r)$. 
The static potential $\tilde V_{pot}(q)$ is computed in 
perturbation theory 
\ba
\label{stapot}
\tilde V_{pot}(q)=-C_F\frac{\al_s}{q^2}\left(1+\al_s(b_1\ln \mu/q+a_1)
+\al_s^2(b_1^2\ln^2 \mu/q +b_2\ln \mu/q+a_2)+\ldots\right)\, . 
\ea
Here $b_1=2\beta_0$, $b_2=2(\beta_1+2\beta_0 a_1)$,
\be
(4\pi)\beta_0=\frac{11}{3}C_A-\frac{4}{3}T_Fn_f=\frac{23}{3}
\ee
is the first coefficient of the $\beta$-function, 
\be
(4\pi)^2\beta_1={34\over 3}C_A^2-{20\over 3}C_AT_Fn_f-4C_FT_Fn_f
=\frac{116}{3}
\ee
is the second coefficient of the $\beta$-function.
The static potential can be written in the form
\be
\tilde V_{pot}(q)=-C_F\frac{\al_V(q)}{q^2}
\ee
that gives a definition of 
the effective charge $\al_V$ related to the $\MSsch$-scheme coupling
constant
\be
\al_V(\mu)=\al_s(\mu)(1+a_1 \al_s(\mu)+a_2 \al_s(\mu)^2)\, .
\ee
Coefficients $a_{1,2}$ are known. The numerical value for $a_1$ reads 
\cite{Fish}
\[
a_1=\frac{1}{4\pi}\left({31\over 9}C_A-{20\over 9}T_Fn_f\right)
=\frac{1}{4\pi}\left(\frac{43}{9}\right) \, .
\]
The coefficient $a_2$ is given by \cite{Peter,Schr}
\[
(4\pi)^2a_2= \left({4343\over 162}+4\pi^2-{\pi^4\over 4}
+{22\over3}\zeta(3)\right)C_A^2-
\left({1798\over 81} + {56\over 3}\zeta(3)\right)C_AT_Fn_f
\]
\[
-\left({55\over 3} - 16\zeta(3)\right)C_FT_Fn_f
+\left({20\over 9}T_Fn_f\right)^2\, ,
\]
or in a more concise form,
\[
a_2= \frac{1}{(4\pi)^2}\left(\frac{7217}{162} + 36\pi^2 - \frac{9\pi^4}{4} - 
\frac{62\zeta(3)}{3}\right)=\frac{1}{(4\pi)^2}155.842...
\]
In QCD we have $C_A=3$, $C_F=4/3$, $T_F=1/2$, and $n_f=5$
for the energy region of the $t\bar t$-pair production. 
The effective coupling 
$\al_V$ is nothing but a running coupling constant in some special
subtraction scheme.
The coefficient $a_2$ allows one to find the effective $\beta$-function 
$\beta_V$ for the evolution of the coupling $\al_V$ in the NNLO.  
Therefore the effective coupling constant for the static potential 
is now fully determined at the NNLO.
\item 
High energy coefficient $C_h(\al_s)$ is given by the expression
\be
\label{hic}
C_h(\al)=1-4\frac{\al_s}{\pi}+C_F\left(\frac{\al_s}{\pi}\right)^2
\left(-\pi^2 \left(\frac{2C_F}{3}+C_A\right)
\ln\frac{\mu_f}{m_t}+c_2\right)\, .
\ee
In NNLO, there appears a new term proportional to the logarithm
of the factorization parameter $\mu_f$
that separates long and short distances 
(or large and small momenta) within the effective theory approach.
The finite ($\mu_f$ independent) 
coefficient $c_2$ is known \cite{coef1,coef2}
\[
c_2=\left({39\over 4}-\zeta(3)+{4\pi^2\over 3}
\ln{2}-{35\pi^2\over 18}\right)C_F-\left({151\over 36}+{13\over 2}
\zeta(3)+{8\pi^2\over 3} \ln{2}-{179\pi^2\over 72}\right)C_A
\]
\[
+\left({44\over 9}-{4\pi^2\over 9}+{11\over 9}n_f\right)T_F .
\]
The coefficient of NLO in eq.~(\ref{hic}) 
is $\mu_f$ independent -- the factorization procedure 
(a separation of scales) 
is insensitive to the border. 
An explicit dependence of high and low energy quantities
on the factorization scale $\mu_f$ 
is a general feature of effective theories which are  
valid only for a given region of energy.
A physical quantity, which is given 
by a proper combination of results obtained 
in different energy regions within respective 
effective theories, is 
factorization scale independent
e.g. \cite{pivKK,hoa:match}.
In NRQCD this feature reveals itself
in $\mu_f$ dependence of Green's function and 
of the high energy coefficient $C_h$.
To see how this dependence  
emerges one can consider the contribution of a generic 
NNLO (three-loop) diagram into the cross section. 
In full QCD one can in principle compute it for an arbitrary 
numerical value of the invariant $s=(2m_t+E)^2$
but it is a rather complicated function of the ratio $s/m_t^2$.
It is difficult to obtain such a function analytically
(numerical study was done in \cite{schw3}).
However, the analytical result for the diagram
in the threshold limit simplifies
and leads to a logarithmic singularity in energy $E$ 
(for the moment we forget about Coulomb singularities)  
\be
\al_s^2\ln(E/m_t) \, , \quad E\to 0 \, .
\ee
Within the effective theory approach this singularity splits 
between GF (the low energy singularity 
$\ln(E/\mu_f)$) and the high energy coefficient 
$C_h(\al_s)$ (the mass singularity $\ln(\mu_f/m_t)$) 
\be
\al_s^2\ln(E/m_t)=\al_s^2\left(\ln(E/\mu_f)+\ln(\mu_f/m_t)\right)\, .
\ee
In both limiting cases ($E=0$, or calculation at threshold,
and calculation within low-energy theory for the GF) 
the resulting integrals are simpler than the
original integral for the diagram and can be done analytically.
For one massive particle the calculation of diagrams is 
technically simpler and the complete 
expressions in full theory are available for some physical quantities
\cite{barH}. The corresponding 
effective theory near the static limit is HQET
which is also simpler than NRQCD.
In this case one can 
analyze the cancellation of factorization scale dependence 
between matching (high energy) coefficients 
\cite{barmatch} and the low energy (HQET) amplitudes explicitly.
\end{itemize}
Collecting the above ingredients together 
allows one to have NNLO accuracy for $\Pi(E)$ near the threshold.

\noindent{\bf 4. Solution for the low energy part.}

\noindent The main dynamical object used to 
describe the $t\bar t$ system near the threshold
is the nonrelativistic Green's function
$G=(H-E)^{-1}$.
The Hamiltonian is represented in the form \cite{AshBer,Landau,anh}
\be
H=H_C+\Delta H, \quad H_C=\frac{p^2}{m}-C_F\frac{\al_s}{r}
\ee
with 
\be
\label{dH}
\Delta H = \Delta V_{pot}-\frac{H_C^2}{4m_t}
-\frac{3C_F\al_s}{4m}\left[H_0,\frac{1}{r}\right]_{+}
-\frac{4\pi\al_s}{m_t^2}
\left(\frac{C_F}{3}+\frac{C_A}{2}\right)\delta(\vec{r}) \, .
\ee
Constructing the Green's function is straightforward and can be done
analytically within perturbation theory near Coulomb Green's 
function $G_C(E)$ or numerically (for complex values of $E$ only 
that can be used 
to describe the production of particles with nonzero width)
\cite{KPP,Hoang,PPres,Mel,BSS1,Nag,direg,Yak,HoaTeu}.
Results are presented basically as an expansion
in consecutive orders
\be
G=G_0+\Delta G_1 +\Delta G_2
\ee
to check the convergence of the approximations.
Here:
\begin{itemize}
\item LO: Coulomb approximation, $G_0=G_C$
\item NLO: $\Delta G_1\to O(\al_s)$ corrections from 
the static potential $V_{pot}(r)$
\item NNLO: $\Delta G_2\to O(\al_s^2)$ corrections from $\al_s^2$
terms in the static potential $V_{pot}(r)$
and from the second iteration of the $O(\al_s)$ term 
in $V_{pot}(r)$, relativistic $v^2$ corrections.
\end{itemize}
\noindent The relativistic $H_C^2$ and anticommutator corrections 
in eq.~(\ref{dH}) can be taken
into account by the shift of the parameters 
of the Coulomb Green's function $E\to E+E^2/4m_t$
and $\al_s\to \al_s(1+3E/2m_t)$. In this respect the modified Coulomb
approximation can be used as the leading order approximation. 
The analytical solution for Green's function is perturbative 
\be
G=G_C-G_C \Delta H G_C +G_C \Delta H G_C \Delta H G_C-\ldots
\ee
The $\MSsch$-scheme for the static potential
$V_{pot}(r)$ has been used in the solution.
The numerical results obtained by the different authors are 
rather close to each other \cite{revHoa}.

\noindent{\bf 5. Features of the physical result.}

\noindent The top quark width $\G_t$ plays a crucial role
in the calculation of the $t\bar t$ production
cross section near the threshold \cite{FK}.
At the calculational level 
the width can be taken into account by a shift of the energy variable
$E$. Operationally one can proceed as follows. 
The mass operator of the top quark is 
approximated by the expression 
$M=m_t-i\G_t/2$. Then the kinematical variable $s-4m_t^2$
relevant to the near-threshold dynamics
is substituted with $s-4M^2$ ($\sqrt{s}=E+2m_t$)
and one finds 
\[
s-4M^2=4m_t(E+i\G_t)+E^2+\G_t^2\, .
\]
Neglecting higher orders in $E$ and $\G_t$ one obtains a recipe
for taking into account the width $\G_t$ by the shift $E\to E+i\G_t$.
The dispersion relation for the vacuum polarization function $\Pi(E)$
has the form 
\[
\Pi (E)=\int \frac{\rho(E')dE'}{E'-E}\, .
\]
With the shift recipe one finds 
\be
\label{dispr}
\sigma(E)\sim {\rm Im}~\Pi (E+i\G_t)
= {\rm Im}~\int \frac{\rho(E')dE'}{E'-E-i\G_t}
= \G_t\int \frac{\rho(E')dE'}{(E'-E)^2+\G_t^2}\, .
\ee
Because the point $E+i\G_t$ lies sufficiently far from the positive semiaxis
(and the origin) in the complex energy plane
the cross section eq.~(\ref{dispr}) is calculable point-wise in 
energy.
For the $b\bar b$ system where $\G_b$ is small the situation is
different and only moments of different kinds, for instance, 
\[
\int_{4m_b^2}^\infty \frac{\rho(s)ds}{s^n}
\]
are meaningful in the near-threshold Coulomb
PT calculations.
The hadronic cross section $\sigma(E)$ was obtained by many authors
(as a review see, \cite{revHoa}).
\begin{figure}[!ht]
       \epsfig{file=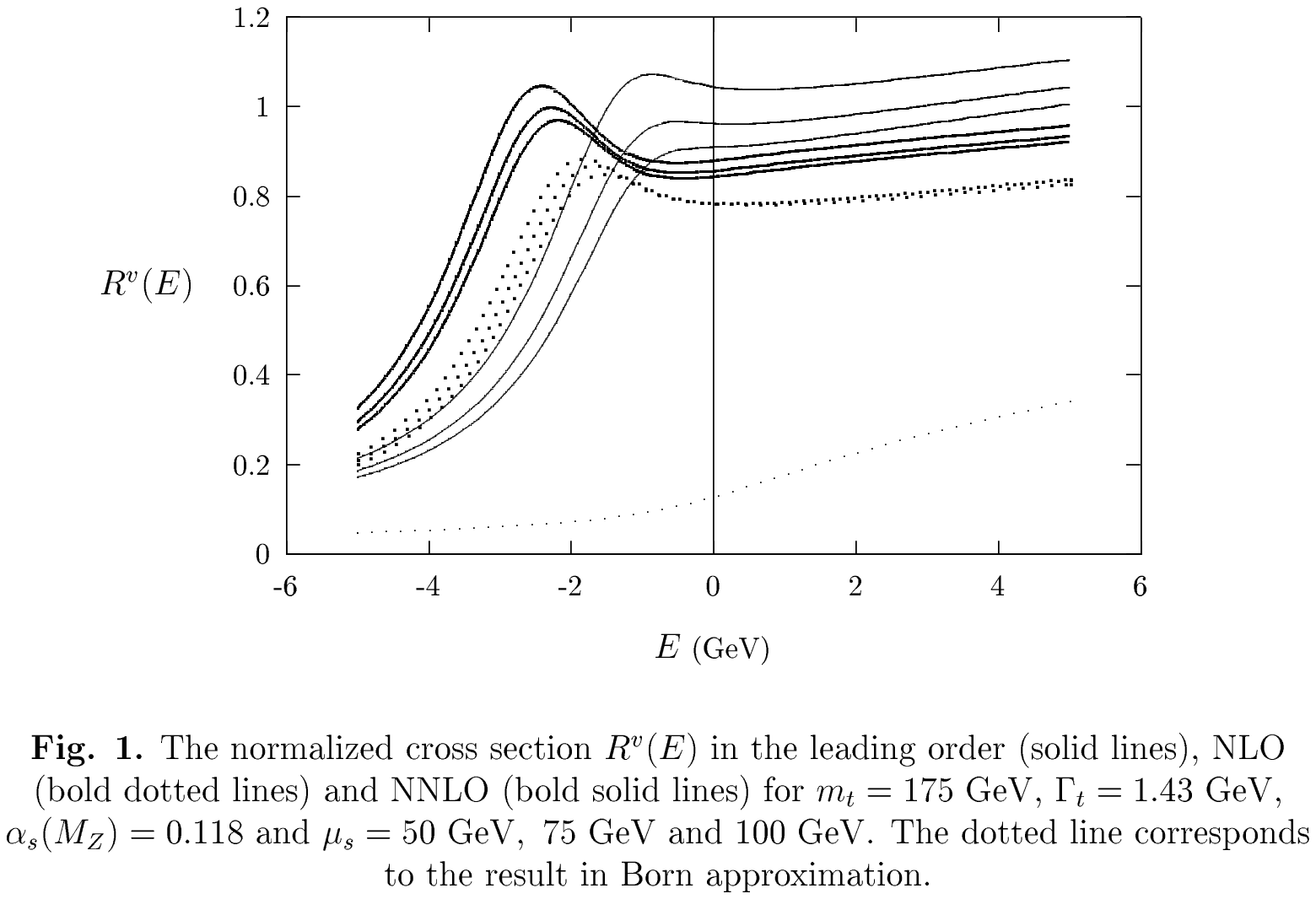,%
angle=0,%
width=1.\textwidth,%
height=0.5\textheight%
}
\end{figure}
The normalized cross sections $R^v(E)$
for typical values $m_t=175$~GeV, $\G_t=1.43$~GeV, 
$\al_s(M_Z) = 0.118$ are plotted in Fig.1 \cite{direg}.
The curves have characteristic points
which are usually considered as basic observables.
They are:
$E_p$ -- the position of the peak in the cross section
and $H_p$ -- the height of the peak in the cross section.
In the limit of the small $\G_t$ (at least for $\G_t$
which is smaller than the spacing between 
the first two resonances) one would have 
$E_p\sim E_0$ and $H_p\sim |\psi_0(0)|^2$.
For the actual value of $\G_t=1.43$~GeV which is larger than the 
spacing being the first two resonances 
the peak position and height are not determined by the first resonance
only. 
The convergence for $E_p$ and $H_p$ in consecutive orders 
of perturbation theory near the Coulomb solution 
is not fast in the $\MSsch$-scheme. 
For the typical numerical values of the theoretical parameters 
$m_t$, $\G_t$ and $\al_s(M_Z)$ one finds \cite{direg}
\ba
E_p&=&E_0(1+0.58+0.38+\ldots)\nn \\
H_p&=&H_0(1-0.15+0.12+\ldots)
\ea 
(see also \cite{PY,Y1}).
Important contributions that affect the quality of
convergence are the local 
term ($\sim \al_s V_0\delta(\vec{r})$ which is related to $1/r^2$
non-Abelian term \cite{Gup})
and higher order PT corrections to $V_{pot}(r)$.

\noindent{\bf 6. Possible improvements.}

\noindent The slow convergence
for the peak characteristics of the cross section
has caused some discussion.
The suggestions of the redefinition of the top quark mass
have been made (as a review see \cite{revHoa}, also \cite{recren,YakM}).
The use of the pole mass as a theoretical parameter for the
description of the cross section near the threshold is criticized 
on the ground of its infrared instability \cite{mpoleIR}.
It is usual that some effects of interaction can be taken
into account by introducing the effective mass parameter
for the particle \cite{polaron,politz,effq}.
In this talk I only discuss some possible ways of
optimizing the convergence for the Green's function
with the pole mass as a theoretical parameter.
Note that actual 
calculations near the threshold have been performed
within the pole mass scheme. For optimizing the convergence
one can use methods of exact summation of some contributions
in all orders and renormalization scheme invariance 
of PT series e.g. \cite{KPP}.
The Hamiltonian can be written in the form 
\be
H=H_{LO}+\Delta V_{PT}+ \al_s V_0\delta(\vec{r}), 
\quad V_0=-\frac{4\pi}{m_t^2}\left(\frac{C_F}{3}+\frac{C_A}{2}\right)
\ee
where corrections are given by 
the perturbation theory corrections to $V_{pot}(r)$ 
($\Delta V_{PT}$-part) and by the local term
($\al_s V_0\delta(\vec{r})$-part).
These two contributions can be dealt with more
accurately than in the straightforward approach.
Indeed,
the $\delta(\vec{r})$-part is a separable potential and can be taken 
into account exactly \cite{nonrelPiv}.
The solution reads
\be
G(E;0,0)=\frac{G_{ir}(E;0,0)}{1+\al_s V_0 G_{ir}(E;0,0)}
\ee
with 
\be
G_{ir}(E)=(H_{LO}+\Delta V_{PT}-E)^{-1} 
\ee
being the irreducible Green's function.
Dealing with the PT expansion of the static potential 
in NRQCD is an important issue in getting
stable results for the cross section near the threshold 
because the static potential 
is the genuine quantity which is
computed in high order of PT in the strong coupling constant
\cite{av}. 
The convergence in the $\MSsch$ scheme 
is not fast which reflects the
physical situation that the observables 
represented by the cross section curve (for instance,
$E_p$ and $H_p$) are sensitive to
different scales. The finite-order perturbation 
theory expansion of the static potential 
given in eq.~(\ref{stapot})
cannot handle several distinct scales with the
same accuracy. Indeed, the PT expansion of the static potential 
is done near some (arbitrary) scale (or distance) 
which can be considered simply as a normalization point.
The farther a given point lies from this normalization point
the worse the precision of the PT expansion 
for the static potential at this point is.
The PT expressions in the $\MSsch$ scheme 
are not directly sensitive to physical scales 
because subtractions are made
in a mass independent way (for instance, 
massive particles do not decouple
automatically in the $\MSsch$ scheme e.g. \cite{decoupl0}). 
Therefore it is instructive to rewrite 
the static potential in more physical terms than just 
the $\MSsch$-scheme parameters $(\al_{\MSsch},\mu)$ 
(which would remind the reader of the momentum subtraction
scheme where, for instance, the decoupling is explicit \cite{decoup})
\ba
V_{pot}(r)&=&
-C_F\frac{\al_0}{r}\left(1+\al_0 b_1\ln r/r_0
+\al_0^2 (b_1^2\ln^2 r/r_0 +b_2'\ln r/r_0+c)+\ldots\right)  \, .
\ea
Here $r_0$ and $c$ are the parameters of 
the renormalization scheme freedom in
NNLO and $\al_0$ is the corresponding coupling in the
$\{r_0,c\}$-scheme \cite{stevenson}.
They parameterize 
the center of the expansion (a normalization point)
and the derivative (respective $\beta$-function)
of the static potential. The parameters ($r_0,c$) can be chosen such 
in order to minimize the higher order corrections to a particular 
observable (e.g. \cite{KPP}
where NLO analysis has been done).
In such a case $r_0$ can be understood 
as a typical distance to which a chosen observable is sensitive. 
Note that the best approximation of the static potential $V_{pot}(r)$
for different scales would be provided by the use of 
the running coupling constant $\al_s(r)$.
The analytical calculation of the Green's function
becomes just impossible in this case. However,   
even for the numerical calculation of the Green's function
one cannot naively use the running coupling constant 
$\al_s(r)$ in the static potential $V_{pot}(r)$
for all $r$. With a generic running coupling constant, 
the IR singularity (Landau pole) can occur
in $\al_s(r)$ at large $r$ which are 
formally necessary for computing the Green's function.
This singularity has little to do with usual divergences 
within the effective theory approach and cannot be removed by the
standard renormalization tools.
It can be dealt with if an IR fixed point appears in the evolution
for the effective coupling constant (e.g. \cite{renorm})
or with some other regularization e.g. \cite{richard}.
For the top quark production the contribution of this area (large $r$)
into the cross section is small because of the large decay width
of the top quark.
In the finite-order PT analysis the parameters 
$r_0$ and $c$ can be chosen to minimize higher order corrections
either to $E_p$ or to $H_p$ but not to both simultaneously
because $E_p$ and $H_p$ are sensitive to different distances.
Indeed, one finds the difference of scales minimizing corrections 
to the first Coulomb resonance in NLO to be
\be
\label{relsc}
\ln(r_{E}/r_\psi)=\frac{1}{3}+\frac{\pi^2}{9} \, .
\ee
Because of the large top quark width 
many states (resonances and continuum alike) contribute
into the position and height of the peak in the cross section.
Therefore the characteristic distance estimates are 
not so transparent (the NNLO peak position, for instance, is not exactly
the ground state energy in zero width limit). 
The relation (\ref{relsc}) can serve just as a basic guide.
In practical analysis one can choose the particular numerical values for 
the parameters ($r_0,c$) which stabilize either $E_p$ or $H_p$. 

\noindent{\bf 7. Conclusion}

\noindent
To conclude, we have presented a cross section for 
the $t\bar t$ pair production near the threshold.
The result is based on the solution to the
Schr{\"o}dinger equation for the nonrelativistic Green's function.
Some methods of resummation and convergence optimization 
are discussed.

\vspace{5mm}
\noindent
{\large \bf Acknowledgments}\\[2mm]
This work is partially supported
by Volkswagen Foundation under contract
No.~I/73611 and Russian Fund for Basic Research under contract
99-01-00091.


\begin{thebibliography}{99}
\bibitem{exp}E.Accomando {\it et al.}, Phys.Rep. {\bf 299}(1998)1.
\bibitem{CasLep1} W.E.Caswell and G.E.Lepage, Phys.Lett.
                 {\bf B167}(1986)437.
\bibitem{CasLep2}G.E.Lepage {\it et al.},  Phys.Rev. {\bf D46}(1992)4052
\bibitem{CasLep3}G.T.Bodwin, E.Braaten and G.P.Lepage,
                 Phys.Rev. {\bf D51}(1995)1125.           
\bibitem{BetheSal}E.E.Salpeter, H.A.Bethe, Phys.Rev. {\bf 84}(1951)1232.
\bibitem{VL} M.B.Voloshin, Nucl.Phys. {\bf B154}(1979)365;\\
             H.Leutwyler, Phys.Lett. {\bf B98}(1981)447.
\bibitem{ttee} W.Kwong, Phys.Rev. {\bf D43}(1991)1488;\\
               M.J.Strassler and M.E.Peskin,  Phys.Rev.
               {\bf D43}(1991)1500.
\bibitem{FK}   V.S.Fadin and V.A.Khoze, Pis'ma Zh.Eksp.Teor.Fiz.
               {\bf 46}(1987)417; \\ Yad.Fiz. {\bf 48}(1988)487.
\bibitem{PDG}C.Caso {\it et al.}, Eur.Phys.J. {\bf C3}(1998)1. 
\bibitem{ttee1} M.Jezabek,  J.H.K{\"u}hn and T.Teubner, Z.Phys.
               {\bf C56}(1992)653;\\
               Y.Sumino {\it et al.},  Phys.Rev. {\bf D47}(1993)56.
\bibitem{ttgg} V.S.Fadin and V.A.Khoze, Yad.Fiz. {\bf 93}(1991)1118;\\
               I.I.Bigi, V.S.Fadin  and V.A.Khoze, Nucl.Phys. {\bf
               B377}(1992)461;\\
               I.I.Bigi, F.Gabbiani and V.A.Khoze, Nucl.Phys.
               {\bf B406}(1993)3.
\bibitem{ttgg1} J.H.K\"uhn, E.Mirkes and J.Steegborn, Z.Phys. 
               {\bf C57}(1993)615. 
\bibitem{KMgg}  B.Kamal, Z.Merebashvili and A.P. Contogouris,  
                Phys.Rev. {\bf D51}(1995)4808;\\  
                Phys.Rev. {\bf D55}(1997)3229 (Erratum).
\bibitem{Piv}A.A.Pivovarov, Phys.Rev. {\bf D47}(1993)5183.
\bibitem{PP2}A.A.Penin and A.A.Pivovarov, Nucl.Phys. {\bf B550}(1999)375.
\bibitem{INOK} K.A.Ispiryan {\it et al.}, Yad.Fiz. {\bf 11}(1970)1278.
\bibitem{Tar}R.Tarrach, Nucl.Phys. {\bf B183}(1981)384.
\bibitem{Kar}R.Karplus and A.Klein, Phys.Rev. {\bf 87}(1952)848;\\
                G.K{\"a}llen and A.Sarby,
                K.Dan.Vidensk.Selsk.Mat.-Fis.Medd. {\bf 29}(1955), N17, 1;\\
                R.Barbieri {\it et al.}, Phys.Lett. {\bf B57}(1975)535.
\bibitem{HarBr}  I.Harris and L.M.Brown, Phys.Rev. {\bf 105}(1957)1656;\\
                 R.Barbieri {\it et al.}, Nucl.Phys. {\bf B154}(1979)535.
\bibitem{Chi} B.M.Chibisov and M.V.Voloshin, 
Mod.Phys.Lett. {\bf A13}(1998)973.
\bibitem{Fish}   W.Fisher, Nucl.Phys. {\bf B129}(1977)157;\\
                 A.Billoire, Phys.Lett. {\bf B92}(1980)343.
\bibitem{Peter}M.Peter, Phys.Rev.Lett. {\bf 78}(1997)602; 
Nucl.Phys {\bf B501}(1997)471.
\bibitem{Schr}Y.Schr{\"o}der,  Phys.Lett. {\bf B447}(1999)321.
\bibitem{coef1}A.Czarnecky and K.Melnikov,
              Phys.Rev.Lett. {\bf 80}(1998)2531.
\bibitem{coef2}M.Beneke, A.Signer and V.A.Smirnov,
              Phys.Rev.Lett. {\bf 80}(1998)2535.
\bibitem{pivKK}A.A.Pivovarov, JETP Lett. {\bf 53}(1991)536;\\
Phys.Lett. {\bf B236}(1990)214, Phys.Lett. {\bf B263}(1991)282.
\bibitem{hoa:match}  A.H.Hoang, Phys.Rev. {\bf D56}(1997)7276.
\bibitem{schw3}K.G.Chetyrkin, J.H.Kuhn, M.Steinhauser, \\
Phys.Lett. {\bf B371}(1996)93,  Nucl.Phys. {\bf B482}(1996)213.
\bibitem{barH}S.Groote, J.G.K\"orner, A.A.Pivovarov,\\
Phys.Rev. {\bf D61}(2000)071501 [hep-ph/9911393].
\bibitem{barmatch}A.G.Grozin and O.I.Yakovlev,
Phys.Lett. {\bf 285B}(1992)254.
\bibitem{AshBer}A.I.Achieser and V.B.Berestezki,
Quantum electrodynamics, Moscow 1959.
\bibitem{Landau} L.D.Landau and E.M.Lifshitz, Relativistic Quantum
                 Theory, Part 1 (Pergamon, Oxford, 1974).
\bibitem{anh}I.G.Halliday, P.Suranyi,
Phys.Rev. {\bf D21}(1980)1529. 
\bibitem{KPP}  J.H.K{\"u}hn, A.A.Penin and A.A.Pivovarov,
               Nucl.Phys. {\bf B534}(1998)356.
\bibitem{Hoang}A.H.Hoang and T.Teubner,
Phys.Rev. {\bf D58}(1998)114023.
\bibitem{PPres}A.A.Penin and A.A.Pivovarov, Phys.Lett. {\bf B435}(1998)413;
Phys.Lett. {\bf B443}(1998)264; Nucl.Phys. {\bf B549}(1999)217.
\bibitem{Mel}K.Melnikov and A.Yelkhovsky, Nucl.Phys. {\bf B528}(1998)59.
\bibitem{BSS1}M.Beneke, A.Singer and V.A.Smirnov,
              Phys.Lett. {\bf B454}(1999)137.      
\bibitem{Nag}T.Nagano, A.Ota and  Y.Sumino, 
               Phys.Rev. {\bf D60}(1999)114014.
\bibitem{direg}A.A.Penin and A.A.Pivovarov, 
MZ-TH-98-61, Dec 1998. 41pp. [hep-ph/9904278],\\
to be published in Yad.Fiz. 
\bibitem{Yak}O.Yakovlev, Phys.Lett. {\bf B457}(1999)170.
\bibitem{HoaTeu} A.H.Hoang and T.Teubner,   
Phys.Rev. {\bf D60}(1999)114027.
\bibitem{revHoa}A.H.Hoang {\it et al.}, Eur.Phys.J.direct  
{\bf C3}(2000)1 [hep-ph/0001286].
\bibitem{PY}A.Pineda and F.J.Yndurain, Phys.Rev. {\bf D58}(1998)094022.
\bibitem{Y1}F.J.Yndurain, [hep-ph/00007333].
\bibitem{Gup}S.N.Gupta and S.F.Radford,  Phys.Rev. {\bf D24}(1981)2309;\\
Phys.Rev. {\bf D25}(1982)3430 (Erratum);\\
S.N.Gupta, S.F.Radford and W.W.Repko, Phys.Rev. {\bf D26}(1982)3305.
\bibitem{recren}K.G.Chetyrkin, M.Steinhauser, 
Phys.Rev.Lett. {\bf B83}(1999)4001.
\bibitem{YakM}S.Groote and O.Yakovlev, hep-ph/0008156.
\bibitem{mpoleIR}
M.Beneke and V.M.Braun, Nucl.Phys. {\bf B426}(1994)301;\\ 
I.I.Bigi {\it et al.},  Phys.Rev. {\bf D50}(1994)2234.
\bibitem{polaron}R.P.Feynman, Phys.Rev. {\bf 97}(1955)660.
\bibitem{politz}H.D.Politzer, Nucl.Phys.\ {\bf B117}(1976)397. 
\bibitem{effq}N.V.Krasnikov and A.A.Pivovarov,
Russ.Phys.J. {\bf 25}(1982)55.
\bibitem{nonrelPiv}A.A.Pivovarov, Phys.Lett. {\bf B475}(2000)135.
\bibitem{av}M.Jezabek, J.H.K{\"u}hn, M.Peter, Y.Sumino and
T.Teubner,\\ 
Phys.Rev. {\bf D58}(1998)014006. 
\bibitem{decoupl0}K.G.Chetyrkin, B.A.Kniehl, M.Steinhauser,
Phys.Rev.Lett. {\bf 79}(1997)2184.
\bibitem{decoup}T.Appelquist and J.Carazzone, 
Phys.Rev. {\bf D11}(1975)2856.
\bibitem{stevenson}P.M.Stevenson, Phys.Rev. {\bf D23}(1981)2916.
\bibitem{renorm}A.A.Penin and A.A.Pivovarov,
Phys.Lett. {\bf B401}(1997)294;\\
N.V.Krasnikov and A.A.Pivovarov,
Mod.Phys.Lett. {\bf A11}(1996)835, hep-ph/9510207,
hep-ph/9512213.
\bibitem{richard}J.L.Richardson, Phys.Lett. {\bf B82}(1979)272.

\end{thebibliography}
\end{document}